\documentclass{article}    

\usepackage{url}  
\usepackage{ifthen}  
\usepackage{multicol}   
\usepackage{multirow} 
\usepackage{url}
\usepackage{caption}
\usepackage[latin1]{inputenc} 
\usepackage{graphicx}
\usepackage{changepage}

\title{Testing the AgreementMaker System in the Anatomy Task of OAEI 2012}
 
\author{Daniel Faria$^1$ \and Catia Pesquita$^1$	\and Emanuel Santos$^1$ \\
\and Francisco M. Couto$^1$ \and Cosmin Stroe$^2$ \and Isabel F. Cruz$^2$ \and \\
   $^1$ \small{LaSIGE, Department of Informatics, Faculty of Sciences of the University of Lisbon, Portugal} \\
	 $^2$ \small{ADVIS Lab, Department of Computer Science, University of Illinois at Chicago, USA}}

\begin{document}

\maketitle

\begin{abstract}
The AgreementMaker system was the leading system in the anatomy task of the Ontology Alignment Evaluation Initiative (OAEI) competition in 2011. While AgreementMaker did not compete in OAEI 2012, here we report on its performance in the 2012 anatomy task, using the same configurations of AgreementMaker submitted to OAEI 2011. Additionally, we also test AgreementMaker using an updated version of the UBERON ontology as a mediating ontology, and otherwise identical configurations. \\ AgreementMaker achieved an F-measure of 91.8\% with the 2011 configurations, and an F-measure of 92.2\% with the updated UBERON ontology. Thus, AgreementMaker would have been the second best system had it competed in the anatomy task of OAEI 2012, and only 0.1\% below the F-measure of the best system.
\end{abstract}

\section{Introduction}
AgreementMaker is a powerful, flexible and extensible system for matching ontologies and schemas that has been in development since 2001 \cite{Cruz09}. Originally focused on geospatial applications, AgreementMaker has since been expanded to include many types of matching algorithms and thus handle diverse applications. \\ In addition to its use in practical applications, the AgreementMaker system has been tested in the Ontology Alignment Evaluation Initiative (OAEI) competition. The development of AgreementMaker focused in particular on the anatomy task, which is a real world application consisting in matching the Adult Mouse Anatomy and the NCI Thesaurus describing the human anatomy. In this task, AgreementMaker was the best performing system, both in OAEI 2010 and OAEI 2011. \\ While the AgreementMaker system did not compete in the OAEI 2012, here we report on its performance in the 2012 anatomy task, using the configurations used for OAEI 2011 \cite{Cruz11}. Additionally, since one of the matching algorithms used in OAEI 2011 uses UBERON \cite{Mungall12} as a mediating ontology, we also tested AgreementMaker with an updated version of UBERON.  

\section{Methods}
The AgreementMaker system was configured as submitted to OAEI 2011 \cite{Cruz11}. However, in addition to testing AgreementMaker with the UBERON version used in OAEI 2011 (release 2011-08-31) we also tested it a more recent version of UBERON (release 2012-08-14) \cite{Mungall12}. \\ We tested the AgreementMaker system on the anatomy task of OAEI 2012, using the reference alignment provided by the competition to evaluate its performance.

\section{Results}
The results of the AgreementMaker system in the anatomy task of OAEI 2012 are presented in Table \ref{tab:am}. As expected, the results obtained with the configurations submitted to OAEI 2011 match the results obtained in that competition, given that no changes were made to the ontologies or reference alignment for the task. The results obtained with the updated UBERON ontology had a slightly lower precision, but a 0.9\% higher recall and thus a higher F-measure. \\ In comparison with the systems that participated in OAEI 2012 (Table \ref{tab:oaei}) the recall and F-measure of AgreementMaker are second only to those of the GOMMA-bk matcher \cite{GOMMA}. Furthermore, when using the more recent release of UBERON, the AgreementMaker is below GOMMA-bk by only 0.1\% F-measure, and although its recall is 3.5\% lower, its precision is 3.5\% higher.

\begin{table}[htbp]
\begin{adjustwidth}{-60pt}{-60pt}
\centering
\caption{Results obtained with the AgreementMaker system in the OAEI 2012 anatomy task.}
\label{tab:am}
\begin{tabular}{|c|c|c|c|c|c|c|c|}
\hline
\textbf{Matcher} & \textbf{Runtime} & \textbf{Size} & \textbf{Precision} & \textbf{F-Measure} & \textbf{Recall} & \textbf{Recall+} & \textbf{Coherent} \\ \hline
AgrMkr2011 & N/A & 1404 & 0.954 & 0.918 & 0.884 & N/A & - \\ \hline
AgrMkr2011+ & N/A & 1421 & 0.952 & 0.922 & 0.893 & N/A & - \\ \hline
\end{tabular}
\caption*{AgrMkr2011 corresponds to the configuration submitted to OAEI 2011; AgrMkr2011+ is otherwise identical but uses the more recent release of the UBERON mediating ontology.}
\caption{OAEI 2012 anatomy Results.}
\label{tab:oaei}
\begin{tabular}{|c|c|c|c|c|c|c|c|}
\hline
\textbf{Matcher} & \textbf{Runtime} & \textbf{Size} & \textbf{Precision} & \textbf{F-Measure} & \textbf{Recall} & \textbf{Recall+} & \textbf{Coherent} \\ \hline
GOMMA-bk & 15 & 1534 & 0.917 & 0.923 & 0.928 & 0.813 & - \\ \hline
YAM++ & 69 & 1378 & 0.943 & 0.898 & 0.858 & 0.635 & - \\ \hline
CODI & 880 & 1297 & 0.966 & 0.891 & 0.827 & 0.562 & x \\ \hline
LogMap & 20 & 1392 & 0.92 & 0.881 & 0.845 & 0.593 & x \\ \hline
GOMMA & 17 & 1264 & 0.956 & 0.87 & 0.797 & 0.471 & - \\ \hline
MapSSS & 453 & 1212 & 0.935 & 0.831 & 0.747 & 0.337 & - \\ \hline
WeSeE & 15833 & 1266 & 0.911 & 0.829 & 0.761 & 0.379 & - \\ \hline
LogMapLt & 6 & 1147 & 0.963 & 0.829 & 0.728 & 0.29 & - \\ \hline
TOAST* & 3464 & 1339 & 0.854 & 0.801 & 0.755 & 0.401 & - \\ \hline
ServOMap & 34 & 972 & 0.996 & 0.778 & 0.639 & 0.054 & - \\ \hline
ServOMapLt & 23 & 976 & 0.99 & 0.775 & 0.637 & 0.052 & - \\ \hline
HotMatch & 672 & 989 & 0.979 & 0.773 & 0.639 & 0.145 & - \\ \hline
AROMA & 29 & 1205 & 0.865 & 0.766 & 0.687 & 0.321 & - \\ \hline
StringEquiv & - & 946 & 0.997 & 0.766 & 0.622 & 0 & - \\ \hline
Wmatch & 17130 & 1184 & 0.864 & 0.758 & 0.675 & 0.157 & - \\ \hline
Optima & 6460 & 1038 & 0.854 & 0.694 & 0.584 & 0.133 & - \\ \hline
Hertuda & 317 & 1479 & 0.69 & 0.681 & 0.673 & 0.154 & - \\ \hline
MaasMatch++ & 28890 & 2737 & 0.434 & 0.559 & 0.784 & 0.501 & - \\ \hline
\end{tabular}
\caption*{These results are presented as published by the OAEI \cite{OAEI12}.}
\end{adjustwidth}
\end{table}

\section{Conclusions}
The results produced by AgreementMaker in the anatomy task indicate that, had it competed in OAEI 2012 with the exact same system submitted in the previous year, it would have placed second in this task (ranked by F-measure). Furthermore, with just an update to the UBERON ontology used as a mediator, the performance of AgreementMaker would have been only marginally below that of the top system, GOMMA-bk. Thus, although the development of AgreementMaker did not focus on the anatomy task in the past year, the system is still one of the best in this task. \\ The future development of the AgreementMaker system should focus particularly on improving the recall, given that it was significantly below GOMMA-bk in this regard. Strategies for this may include the use of additional mediating ontologies or other external data sources such as the Wikipedia. \\ The development of AgreementMaker should also focus on efficiency, as multiple systems ran the anatomy task in under a minute, whereas the running time of AgreementMaker in OAEI 2011 was 10 minutes. \\ Finally, the development of AgreementMaker should focus on alignment coherency, which is an aspect under evaluation in the OAEI since 2011, but which is currently not addressed by AgreementMaker. Thus, adding a coherence analysis module to the AgreementMaker system is paramount.

\section{Acknowledgments}
DF, CP, ES and FMC were funded by the Portuguese FCT through the SOMER project (PTDC/EIA-EIA/119119/2010) and the multi-annual funding program to LASIGE. CS and IFC were funded by NSF Awards IIS-0812258, IIS-1143926, and IIS-1213013.

\bibliographystyle{plos}  
\bibliography{references}     

\end{document}